\documentclass{revtex4}
\pdfoutput=1
\usepackage{graphicx}
\usepackage{verbatim}  

\def\be{\begin{equation}} 
\def\ee{\end{equation}} 
\def\bea{\begin{eqnarray}} 
\def\eea{\end{eqnarray}} 

\begin{document}

\title{The High-Redshift Neutral Hydrogen Signature of an Anisotropic Matter Power Spectrum}

\author{Oscar F. Hernandez$^{2,1}$}
\author{Gilbert P. Holder$^1$}
\affiliation{$^1$Physics Department, McGill University, 3600 rue University,
Montreal, Quebec, Canada H3A 2T8} 
\affiliation{$^2$Marianopolis College, 4873 Westmount Ave, Westmount, Quebec 
Canada H3Y 1X9}

\begin{abstract}
An anisotropic power spectrum will have a clear signature in the 21cm radiation
from high-redshift hydrogen. We calculate the expected power spectrum of
the intensity fluctuations in neutral hydrogen from before the epoch
of reionization, and predict the accuracy to which future experiments
could constrain a quadrupole anisotropy in the power spectrum. We find
that the Square Kilometer Array  
will have marginal detection abilities for this signal at $z \sim 17$
if the process of reionization has not yet started;
reionization could enhance the detectability substantially. 
Pushing to higher redshifts and higher sensitivity will allow highly
precise (percent level) measurements of anisotropy.
\end{abstract}
\maketitle

\section{Introduction}

Modern cosmology assumes that the universe is both homogeneous and isotropic when averaged on scales larger than 100 Mpc. Observations, in particular that of the cosmic microwave background (CMB), have established homogeneity and isotropy as valid assumptions if we interpret them statistically. Put another way, ``the density of matter may differ from one point in the Universe to another, but the distribution of matter is described as a realization of a random field with a variance that is everywhere the same and the same in every direction''~\cite{Pullen:2007tu}. With the accumulating precision in the measurement of the CMB temperature field and the soon expected polarization data from the Planck satellite, these assumptions can be tested.
Ackerman, Carroll and Wise~\cite{Ackerman:2007nb} have considered the possibility that rotational invariance was broken in a scale invariant way during inflation, by an effect that has since disappeared. 
Their paper suggests a possible explanation for hints of anomalies in 
recent CMB measurements, many of which are carefully discussed,
and dismissed as statistical fluctuations in a recent analysis by the
WMAP team \cite{Bennett:2010} . Such
a preferred direction would be an important clue to the physics of the
early universe.
Pullen and Kamionkowski~\cite{Pullen:2007tu} have developed CMB power spectrum statistics to detect a direction dependence in the fluctuations in temperature or polarization. Groeneboom et al.~\cite{Groeneboom:2009cb} have revisited
the model of Ackerman et al. to include polarization and various 
systematic effects.

While the CMB has been a tremendous resource for exploring the physics
of the early universe, we are approaching the limits of the information
that is encoded: the temperature fluctuations measurements 
expected from the Planck satellite \cite{Planck:2011}  will be primarily
limited by cosmic variance (limits on our ability to characterize variances
due to the finite number of independent and informative 
measurements we can make) in the entire regime where the primary CMB
fluctuations are largely unconfused by astrophysical foregrounds. 
Comparable measurements in polarization would not add a tremendous amount
of new information, with the important exception of a possible detection
of the signature of gravitational radiation generated in the early universe.

Measurements of large scale structure can in principle provide much more
information about the potential fluctuations; the CMB is largely a
two dimensional surface, while surveys can probe the three dimensional
structure of the universe. 
Pullen and Hirata \cite{Pullen:2010} have used luminous red
galaxies in the Sloan Digital Sky Survey to put the strongest
limits on any possible anisotropy, constraining any possible quadrupole
anisotropy in the power spectrum to be less than $40\%$.

On large scales, the fluctuations are small
enough such that they are expected to evolve according to linear theory, but 
small scales are highly non-linear and difficult to model. The scale at which
this transition occurs defines the smallest scale which can be readily used
for measuring cosmological parameters, which places limits on the number of
independent modes of the fluctuations which can be measured. As structure
grows in the universe, this scale progressively moves to larger scales. 
To explore a large number of modes, it is therefore helpful to make
measurements at high redshift.

A promising technique for measuring the three-dimensional structure of
the universe at high redshift is to measure the 21cm hyperfine transition
of neutral hydrogen. In this paper, we investigate the signature of a
preferred axis for the fluctuation power in this redshifted 21cm emission.

Neutral hydrogen is the dominant form of baryonic matter
in the ``dark ages", i.e. before star formation, hence we expect redshifted 
21cm radiation to reach us
from all directions in the sky. The intensity of this
radiation tells us about the distribution of neutral
hydrogen in the universe, as a function of both angular
coordinates in the sky and redshift. Thus, in contrast
to the CMB, 21cm surveys can probe the three-dimensional
distribution of matter in the universe (see~\cite{Furlanetto:2006jb} for an in-depth review).

As a benchmark experiment, we assume that it will be possible to make
arcminute scale measurements with mK sensitivity in a bandwidth
of order 1 MHz. This is a somewhat optimistic forecast, but not far
beyond what is projected for the Square Kilometer Array \cite{Dewdney:2009}.
It is expected that $\sim 250$ square degrees could be surveyed down to 
frequencies of 70 MHz to a sensitivity of 10 mK, with thousands of
frequency resolution elements in a band.
At these frequency and spatial resolutions, the smallest accessible scales
will be of order a few comoving Mpc. We also consider the benefits of
a next generation experiment, an idealized version of an 
Omniscope \cite{Tegmark:2008au}, that can operate at
slightly lower frequencies (to push to higher redshifts), but has a wide
field of view.
This is a challenging experiment, as the noise rises steeply with lower 
frequency, but requires only a modest expansion in frequency coverage and a
feasible increase in sensitivity. 

In what follows below we focus on the signal from before reionization. This
simplifies the analysis and is robust to the details of how the first stars
formed, but is neglecting a possibly large signal \citep{Furlanetto:2006}. 
The 21cm spin temperature
during reionization could be substantially different from the CMB temperature
in the presence of a UV background, providing a substantial
amount of statistical power \citep{Furlanetto:2009}, especially at the
lower range of redshifts we consider, $z \sim 17$. 
The details of reionization are 
sufficiently uncertain at this point that it is difficult to predict
the utility of this epoch for precise measurements of the matter
power spectrum. It is possible that coupling to Ly-$\alpha$ photons
has already coupled the 21-cm spin temperature to the gas temperature
at $z \sim 17$,
but there are currently no observational constraints on this epoch, and
there is no exhaustive theoretical search of parameter space that has
determined the most likely physical conditions at $z \sim 17$. In the
sense of signal-to-noise, this would be a large boost over what is assumed
in what follows, but it comes at the cost of increased astrophysical 
uncertainty.

\section{The 21 cm brightness temperature and its power spectrum}

We can consider the equation of radiative transfer along 
the line of sight through a hydrogen gas cloud and then compare the intensity of the 21 cm coming through the cloud to a hypothetical ``clear view'' of the 21 cm radiation from the CMB radiation. This difference in intensity is expressed through the brightness temperature~\cite{Furlanetto:2006jb} :
 \begin{eqnarray}
\delta T_b(z) & = & \left[ \frac{9 c^3 A_{10} \hbar (1-Y)}{128 \pi G \nu_{21}^2 k_B m_H H_0 \Omega_m^{1/2}}\right]
\left[\frac{x_{HI}(z) (1+\delta_b(z)) (1 + z)^{1/2}}{1+\frac{\partial v^{pec}_\chi/\partial \chi}{H(z)/(1+z)}+{v^{pec}_\chi\over c}} \right] \, 
 \left[1-\frac{T_\gamma(z)}{T_S}\right]
 \label{Tb1} \\
 & =  &  \left[8.8~{\rm mK}\right] 
\left[\frac{x_{HI}(z) (1+\delta_b(z)) (1 + z)^{1/2}}{1+\frac{\partial v^{pec}_\chi/\partial \chi}{H(z)/(1+z)} +{v^{pec}_\chi\over c}} \right] \, 
 \left[1-\frac{T_\gamma(z)}{T_S}\right]
 \label{Tb2}
\end{eqnarray}
where we have taken the values 
$A_{10} = 2.85 \times 10^{-15}~{\rm s}^{-1}$, $H_0=73~{\rm km~s}^{-1}~{\rm Mpc}^{-1}$, 
$\nu_{21}=1420$~MHz,   $\Omega_b=0.0425$, $\Omega_m=0.26$.

The brightness temperature in eq.~\ref{Tb1} is a function of redshift for a chosen line of sight. Since redshift can be related to the radial distance, we can think of the brightness temperature as a function of $\vec x$. Redshift $z$ is related to the comoving radial coordinate $\chi$ through
\be
\chi =  {c \over a_0} \int^{1}_{a_e \over a_0} {da \over a^2 H(a_0 a)} 
\qquad , \qquad
{a_e\over a_0} = {(1+v^{pec}_\chi/c)\over(1+z)}
\label{chi2freq}
\ee
For $1<z<3000$ we have a matter dominated universe with $H(a)=H_0\sqrt{\Omega_m/a^3}$.  Here and throughout, we assume that the peculiar velocities are non-relativistic.
%

After recombination Compton heating through the residual free electrons keep the kinetic cosmic gas temperature $T_K$ and the spin temperature $T_S$, equal to the CMB temperature $T_\gamma$. As illustrated in ref.~\cite{Furlanetto:2006jb}~fig.~6, the cosmic gas begins to decouple from the CMB at around $z\sim300$. At this point the cosmic gas is dense enough that the  spin temperature $T_S$ is cooled from the CMB temperature to the kinetic cosmic gas temperature $T_K$ by collisions. By $z\sim100$ the cosmic gas begins cooling adiabatically as 
$T_K=0.02~{\rm K}(1+z)^2$. 
As the cosmic gas expands collisions between hydrogen atoms become less 
efficient at cooling the spin temperature. Photon interactions
then slow the cooling of the spin temperature, an it is eventually again driven 
towards the CMB temperature. By the time reionization begins at around 
redshifts of 20, $T_S \sim T_\gamma$, making the 21cm signal relatively
small.  In this our analysis we are particularly interested in 
redshifts $z$, with  $15\le z \le 35$ when $T_S \lesssim T_\gamma$.

We now consider fluctuations in the 21 cm brightness temperature by defining an average of $\delta T_b$ over all angular coordinates and defining $\delta_{21}(\vec x)$ through
\be
\delta T_b(\vec x) \equiv \overline{\delta T_b}(z) (1+\delta_{21}(\vec x))
\ee

Fluctuation in the brightness temperature arise through fluctuations in the baryon density $\delta_b$, the neutral fraction $\delta_{x_{HI}}$, fluctuations in the spin temperature $\delta_{T_S}$, and
the line of sight peculiar velocity gradient $\delta_{\partial v}$.

We are interested in $z$ much after recombination but before reionization. During this time the neutral fraction fluctuations are so small that we can just take $x_{HI}=1$. 

The fluctuations in spin temperature can be related to fluctuations in the CMB temperature $\delta_{T_\gamma}$, the cosmic gas kinetic temperature $\delta_{T_K}$ and the baryonic density fluctuations $\delta_b$. This is because the spin temperature is determined solely by the temperatures $T_\gamma$ and $T_K$ as long as UV scattering is negligible, which is true before reionization. The relationship between spin and kinetic gas temperatures is expressed through the collision coefficients $x_c$ which describe
the rate of scattering among hydrogen atoms and 
electrons~\cite{Zygelman:2005,Furlanetto:2006su}:
\be \label{three2}
\left(1-\frac{T_\gamma(z)}{T_S}\right) ={x_c\over1+x_c} \left(1-\frac{T_\gamma(z)}{T_K} \right)
\; ,
\qquad
x_c  =  x^{eH}_c+x^{HH}_c
\; ,
\;
x^{i}_c  =  \frac{n^{i} \kappa^{iH}_{10}}{A_{10}} \frac{T_{\star}}{T_{\gamma}} \, .
\ee
where $k_B T_{\star} \equiv  \hbar 2\pi \nu_{21}=0.0682$~K. 
Fluctuation in the CMB temperature are so small that they can be ignored. 

Combining the remaining contributions to $\delta_{21}$ we calculate that at linear order
\begin{equation}
\delta_{21}(\vec x) = \beta_b(z) \delta_b(\vec x) +  \beta_{T_K}(z) \delta_{T_K}(\vec x) -  \delta_{\partial v}(\vec x),
\label{d21}
\end{equation}
The coefficients $\beta_b$ and $\beta_{T_K}$ are
\begin{eqnarray}
\beta_b(z) & = & 1 + \frac{1}{1+x_c},
\label{betab} \\
\beta_{T_K}(z) & = & \frac{T_\gamma}{T_K - T_\gamma} + \frac{1}{x_c(1+x_c)} \left( x_c^{\rm eH} \frac{\partial \ln \kappa_{10}^{\rm eH}}{\partial \ln T_K} + x_c^{\rm HH} \frac{\partial \ln \kappa_{10}^{\rm HH}}{\partial \ln T_K} \right),
\label{betaT} 
\end{eqnarray}
Now $\delta_{T_K}$ is related to $\delta_b$ via a proportionality constant that depends only on the redshift distance~\cite{Bharadwaj:2004nr}: 
$
\delta_{T_K}(\vec x)=g(z)\delta_b(\vec x) . 
$
Fig. 2 in ref.~\cite{Bharadwaj:2004nr} gives the proportionality constant $g(z)$ for redshift $z$ between 10 and 1000.
We define
\be
\beta(z) \equiv\beta_b(z)  + g(z) \beta_{T_K}(z)
\ee
and hence $\delta_{21}$ in eq.~\ref{d21} becomes 
\begin{equation}
\delta_{21}(\vec x) =\beta(z) \delta_b(\vec x) -  \delta_{\partial v}(\vec x)
\label{d21ii}
\end{equation}
For $15<z<35$,  $\beta(z)$ is approximately 1.6 to within 3.5\% accuracy. 

The line of sight peculiar velocity gradient 
\be
\delta_{\partial v} \equiv \frac{\partial v^{pec}_\chi/\partial \chi}{H(z)/(1+z)}+{v^{pec}_\chi\over c}
\ee
introduce redshift space distortions. For scales large enough for linear theory to hold, Kaiser~\cite{Kaiser:1987qv} has shown that
$
\tilde{\delta}_{\partial v}(\vec k)= - f (\hat{k}\cdot \hat{r}_s)^2 \tilde{\delta}_{matter}(\vec k) \, .
$
Lahav et al.~\cite{Lahav:1991wc} explain that 
$
f(z) = [\Omega_m (1+z)^3\left({H_0/ H(z)}\right)^{2}]^{0.6}
$
which is essentially unity for the matter dominated universe at the redshifts we are considering. Furthermore we assume that $\tilde\delta_b$ follows $\tilde\delta_{matter}$ for the scales we are considering so that
\be
\tilde{\delta}_{\partial v}(\vec k)= - (\hat{k}\cdot \hat{r}_s)^2 \tilde{\delta}_{b}(\vec k)
\ee

We define the power spectra via
\bea
\langle \tilde{\delta}_{21}(\vec k) \, \tilde{\delta}_{21}(\vec k') \rangle 
& \equiv &
(2 \pi)^3 \delta^{(3)}_{D}(\vec k + \vec k') P_{21}(\vec k),
\\
\langle \tilde{\delta}_{b}(\vec k) \, \tilde{\delta}_{b}(\vec k') \rangle 
& \equiv &
(2 \pi)^3 \delta^{(3)}_{D}(\vec k + \vec k') P_{b}(\vec k),
\eea
By writing a spherical harmonic expansion of the 21 cm fluctuations,
$
{\delta}_{21}(\vec x)\equiv{\delta}_{21}(\chi,\hat x)=\sum_{l,m} a_{lm}(\chi) Y_{lm}(\hat x)
$
we can construct the angular power spectrum
\be
\langle a_{lm}(\chi) a^{\dagger}_{lm}(\chi')\rangle=
(4\pi)^2 i^{(l-l')} \int {d^3k\over(2\pi)^3} j_l(k\chi)j_{l'}(k\chi')Y^{*}_{lm}(\hat k)Y_{l'm'}(\hat k)P_{21}(\vec k)\ee

\section{The angular power spectrum in the flat sky distant observer approximation}

We now calculate the 21 cm angular power spectrum in the flat sky approximation distant observer approximation. We look at a tile of sky at the north pole (in the $\hat{z}$ direction), at an average distance $\chi_0$, thickness $\Delta\chi$, and angular size $\Delta\epsilon\times\Delta\epsilon$. The tile is not too thick, so that the constants in redshift z, such as $g(z)$ and $\beta(z)$ do not vary much and we can treat them as constants. In particular from now on we drop the $z$ dependence and write simply $\beta$. As discussed in the previous section, for the redshifts of interest in this paper  $\beta\approx1.6$.  We mask the $\delta(\vec x)$ so that it is non zero only on that tile. This permits us to extend our integrals beyond the tile's dimensions to simplify their evaluation. Incorporating Kaiser's line of sight approximation for red shift space distortions allows us to write
\be
\tilde\delta_{21} (\vec k)= \left(\beta +(\hat{k}\cdot \hat{r}_s)^2 \right)
\,  
\tilde{\delta}_b(\vec k)
\label{d21db}
\ee
where we take the line of sight $\hat{r}_s=\hat{z}$.

We follow~\cite{White:1997wq} for the definition of the $a$'s, and applying it to our case,
for large $l$ we define a 2-D vector $\vec l=(l\cos\phi_l,l\sin\phi_l)$, such that
\bea
a(\vec l,\chi)
& = & \left(
 \sum^{l}_{m=-l} [(i l)^{-m}\sqrt{{4\pi\over2l+1}{(l+m)!\over(l-m)!}}\quad e^{im\phi_l} a_{lm} ]
 \right)
\\
& \sim & \int d^2\epsilon\; e^{(-i \vec l \cdot \vec\epsilon)} \delta_{21}(\chi,\hat x)
\eea
so that
\be
\langle a(\vec l,\chi) a^{\dagger}(\vec l', \chi')\rangle = 
{2\pi\over \chi^2\chi'^2} \delta^{(2)}_{D}({\vec l \over \chi} + {\vec l' \over \chi'}) 
\int {dk_3\over 2\pi}  \left(\beta+ {k^{2}_3\over k^{2}_3+ l^2/\chi^2} \right)^2 \exp{[i k_3 (\chi+\chi')]} \, P_{b}({\vec l \over \chi}, k_3)
\ee
In order to get rid of the $1/\chi^2$ dependence which prevents us from having a diagonalized correlation matrix in momentum space we will define $b \equiv \chi^2 a(\vec l,\chi)$ and we consider $b$ as a function of $\vec\kappa\equiv\vec l/\chi$ and $\chi$ instead of $\vec l$ and $\chi$. We then Fourier transform in $\chi$ and define a 3-D momentum vector $\vec q=(\vec\kappa,k_3)$
\bea
\langle b(\vec q) b^{\dagger}(\vec q')\rangle
& = & (2 \pi)^3 \delta^{(3)}_{D}(\vec q + \vec q') P_{21}(\vec q)
\\
& = & (2 \pi)^3 \delta^{(3)}_{D}(\vec q + \vec q') \left(\beta +(\hat{q}\cdot \hat{z})^2 \right)^2 P_{b}(\vec q)
\label{bb}
\eea

In doing the scaling by distance to make the signal covariance matrix
diagonal, the noise covariance will become more complicated. However,
the noise covariance matrix is expected to already be 
complicated due to non-trivial foregrounds. 

\section{Forecasting the anisotropies}

We follow Pullen-Kamionkowski~\cite{Pullen:2007tu} and expand the anisotropic $P_b(\vec q)$ in terms of spherical harmonics
\be
P_b(\vec q) = \mathcal{P}(q) [1+\sum_{L,M} g_{L,M} (q) Y_{L,M}(\hat{q})]
\ee
To estimate the constraints possible on the $g_{L,M}$'s, we calculate the Fisher matrix. As in \cite{Pullen:2007tu} we will take the $g_{L,M} (q)$ to be constant in $q$. We will first do the case without noise, and then we will consider the case with noise.

The Fisher matrix is
\be
F={1\over2} {\rm Tr}[C,_{g_{L,M}}C^{-1}(C,_{g_{L,M'}})^{\dagger}C^{-1}]
\label{fisher}
\ee
We will consider a quadrupole anisotropy, i.e. L=2.
Eq.~\ref{bb} is the correlation matrix $C_{\vec q, \vec q'}$ and 
\be
C,_{g_{L,M}}=(2\pi)^3 \delta^{(3)}_{D}(\vec q + \vec q')  \left(\beta+(\hat{q}\cdot \hat{z})^2 \right)^2 \mathcal{P}(q) Y_{L,M}(\hat{q})
\label{CgLM}
\ee
so that
\be
C,_{g_{L,M}}C^{-1}(C,_{g_{L,M}})^{\dagger}C^{-1}
=(2\pi)^3 \delta^{(3)}_{D}(\vec q + \vec q') 
{Y_{L,M}(\hat{q}) Y^{*}_{L,M'}(\hat{q})\over [1+\sum_{M''}g_{L,M''}Y_{L,M''}(\hat{q})]^2}
\ee
Taking the trace of the above means setting $q'=-q$ and integrating with measure ${d^3q\over (2\pi)^3}$. This leads to $(2\pi)^3\delta^{3}_D(0)$ which equals the volume of our thick tile. 
Thus we have
\be
F_{M,M'}={1\over2}(Vol)
\int {d^3q \over(2\pi)^3}
{Y_{L,M}(\hat{q}) Y^{*}_{L,M'}(\hat{q})\over [1+\sum_{M''}g_{L,M''}Y_{L,M''}(\hat{q})]^2}
\ee
The measure $\int d^3q=\int^{\infty}_0 q^2 dq \int d\Omega_{\hat q}$ 
can be regulated with the knowledge that
our volume and resolution are both finite; we approximate the volume as 
a pixelized lattice with a total of $N^3\equiv N_T$ pixels. Thus 
\be
{1\over(2\pi)^3}\int^{\infty}_0 q^2 dq \rightarrow {1\over(2\pi)^3}{2\pi\over L}\sum^{N}_{j=1} \left({2\pi j\over L}\right)^2
\approx\left({1\over L}\right)^3 {N^3\over 3}={ N_T\over3 \; Vol}
\ee
Thus we have
\be
F_{M,M'}={ N_T\over6}\int d\Omega_{\hat q}
{Y_{L,M}(\hat{q}) Y^{*}_{L,M'}(\hat{q})\over [1+\sum_{M''}g_{L,M''}Y_{L,M''}(\hat{q})]^2}
\ee
Since we expect the $g_{L,M''}$ to be small we evaluate the integral with $g_{L,M''}=0$. By the orthonormality of the spherical harmonics we get the Fisher matrix without noise:
\be
[F_{M,M'}]_{\rm noiseless}={N_T\over6}\delta_{M,M'}
\ee
This is trivial to invert. 

\begin{figure}
\includegraphics[height=10cm]{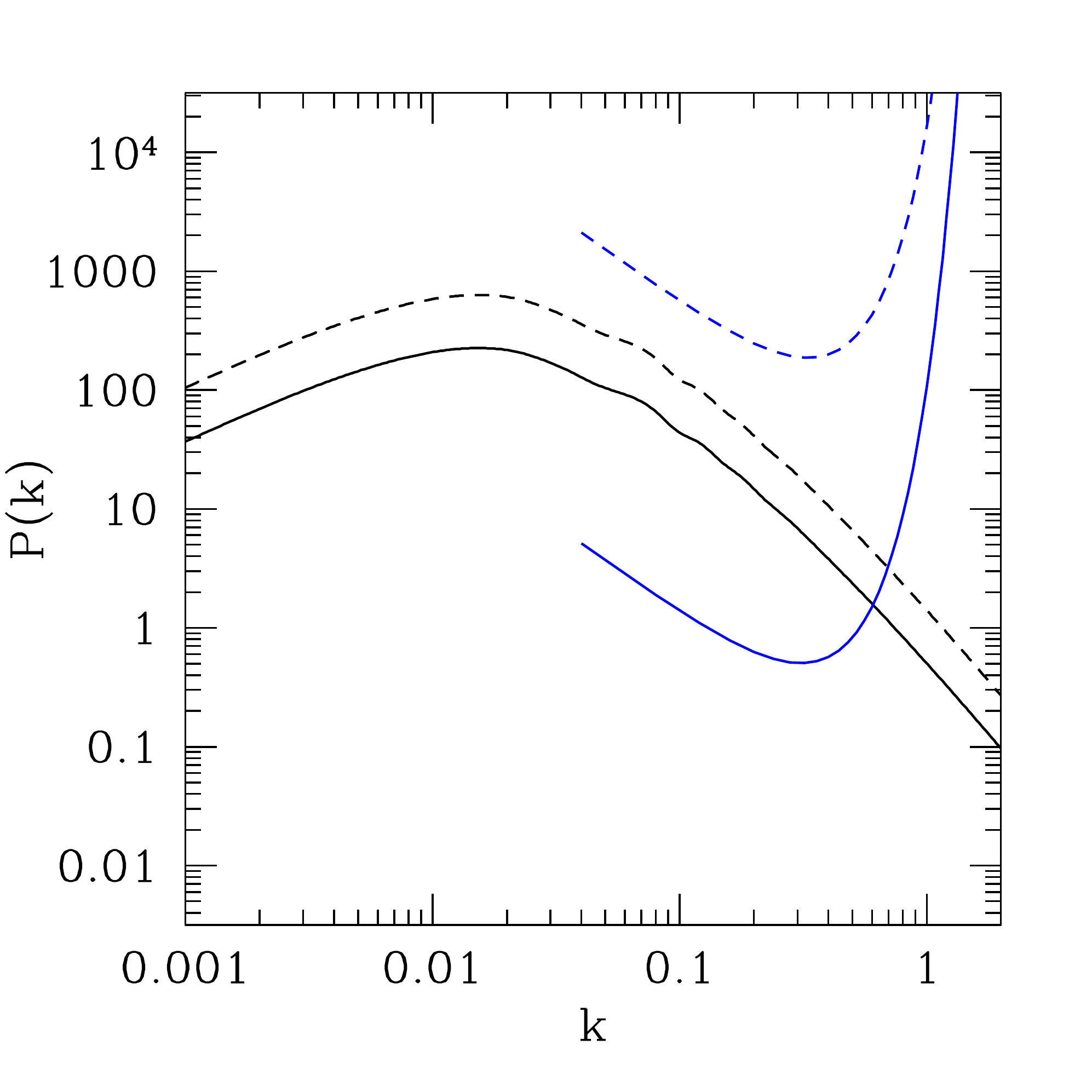}
\caption{Matter power spectra for $z=17.5$ (thick dashed) and $z=30$ 
(thick solid). The lower $z$ curve is higher simply due to the growth
of structure in linear theory. The thin blue curves
show the noise per $ln$ interval in $k$ assuming the experimental 
parameters outlined in the text. These curves include the impact of
both experimental noise and the redshift evolution in the mapping between
density fluctuations and temperature fluctuations. Substantial star 
formation at high $z$ could lower the noise curve at $z\sim17.5$ by up to 
4 orders of magnitude.}
\label{fig:pk}
\end{figure}

Now we include the beam size and instrument noise. 
The correlation matrix given in Eq.~\ref{bb} becomes
\bea
C_{\vec q, \vec q'}
& = & (2 \pi)^3 \delta^{(3)}_{D}(\vec q + \vec q') P_{21}(\vec q)
\\
& = & (2 \pi)^3 \delta^{(3)}_{D}(\vec q + \vec q') 
\left[
\left(\beta +(\hat{q}\cdot \hat{z})^2 \right)^2 P_{b}(\vec q)
\exp(-\sigma_B^2 q^2)
+P_{n}
\right]
\label{corrandnoise}
\eea
This differs from the noiseless case by the addition of two new parameters, $\sigma_B$ and $P_n$. We discuss the meaning of each in turn. 

We assume a Gaussian beam characterized by a beam width $\sigma_B$. This beam width is related to both the radial direction (i.e. frequency resolution as given by Eq.~\ref{chi2freq}) and the transverse direction angular resolution. While in general these two resolutions, $\sigma_\chi$ for the radial and $\sigma_T$ for the transverse, can be different, an experiment well constructed for detecting an anisotropic matter spectrum would choose them so that
$
\sigma_T^2 l^2 + \sigma_\chi^2 k_3^2 = \sigma_B^2 q^2.
$

We model the instrument noise as white, uncorrelated between pixels and uniform in momentum with value $P_n$. We relate the noise per pixel temperature $T_n$ to $P_n$ through
\be
(T_{n}/~\overline{\delta T_b}~)^2=\int {d^3q\over (2 \pi)^3} P_n \exp(-\sigma_{B}^2 q^2/(4\pi))=
{P_n\over \sigma_{B}^3}
\label{noise}
\ee
Here we have incorporated that cutoff as a Gaussian characterized by the beam width,
but any reasonable cutoff in the momentum volume would give similar results. 
Varying the power-law index of the noise (to $1/k$, for example), affects the forecasts. Such considerations can straightforwardly be included in our analysis and we discuss them in the conclusions.  

The derivative of the correlation matrix with respect to the $g_{L=2,M}$'s is 
\be
C,_{g_{L,M}}=(2\pi)^3 \delta^{(3)}_{D}(\vec q + \vec q')  \left(\beta +(\hat{q}\cdot \hat{z})^2 \right)^2 \mathcal{P}(q) \exp(-\sigma_B^2 q^2) Y_{L,M}(\hat{q})
\ee
and the Fisher matrix becomes
\be
F_{M,M'}={1\over2}(Vol)
{1\over(2\pi)^3}\int d^3q
{
Y_{L,M}(\hat{q}) Y^{*}_{L,M'}(\hat{q})
\over 
\left[1+\sum_{M''}g_{L,M''}Y_{L,M''}(\hat{q}) +{P_n \over  \left(\beta +(\hat{q}\cdot \hat{z})^2 \right)^2 \mathcal{P}(q)\exp(-\sigma_B^2 q^2) }\right]^2
}
\ee
Again, since we expect the $g_{L,M''}$ to be small we evaluate the integral with $g_{L,M''}=0$. We are particularly interested in forecasting for the smallest redshifts before ionization, i.e. $15<z<35$ which is where we can take $\beta\approx 1.6$. For these redshifts $\left(\beta +(\hat{q}\cdot \hat{z})^2 \right)^2$ will be between 2.56 and 6.76. We approximate our Fisher matrix by replacing
$(\hat{q}\cdot \hat{z})^2$ by 1/2 so that $\left(\beta +(\hat{q}\cdot \hat{z})^2 \right)^2$ is replaced by 4.41. 
\bea
F_{M,M'}&=&{1\over2}(Vol)\int d\Omega_{\hat q} Y_{L,M}(\hat{q}) Y^{*}_{L,M'}(\hat{q})
{1\over(2\pi)^3}\int_0^{\infty} dq
{
q^2
\over 
\left[1+{P_n \exp(\sigma_B^2 q^2)\over  4.41 \mathcal{P}(q) }\right]^2
}
\\
&=& {1\over2} \delta_{M,M'} (Vol)
{1\over(2\pi)^3}\int_0^{\infty} dq
{
q^2
\over 
\left[1+{\sigma_{B}^3 (T_{n}/~\overline{\delta T_b}~)^2 \exp(\sigma_B^2 q^2)\over  4.41 \mathcal{P}(q) }\right]^2
}
\\
&\equiv& 
{1\over2} \delta_{M,M'} (Vol) ~I 
\label{fishernoise}
\eea
We evaluate numerically the integral  for two cases. Case I will correspond to the Square Kilometer Array target values. Case II will correspond to the Fast Fourier transform telescope (FFTT) described in~\cite{Tegmark:2008au}.

From the thermal noise per visibility as given by Morales~\cite{Morales:2004ca} in eq.~7, we arrive at the thermal noise per pixel for the brightness temperature given by
\be
T_n={\sqrt{2}~T_{\rm sys} \over \sqrt{B~\tau} }{\theta_{\rm diffraction}\over\theta_{\rm desired}}
\label{Tnoise}
\ee
where $T_{\rm sys}$ is the system temperature, $B$ is the bandwidth, and $\tau$ is the total observing time. $\theta_{\rm diffraction}$ is the diffraction limited resolution $\lambda_{21} (1+z) / \sqrt{A_e}$ ($A_e$ is the effective antenna area) and $\theta_{\rm desired}$ is the 1 arcminute resolution we desire. 
The angular resolution is assumed to be tuned by a dilution of the array
from being fully compact by a simple scaling of all baseline positions by
a fixed amount.
The system temperature is given by ARCADE 2~\cite{Fixsen:2009xn} as 
\be
T_{\rm sys} = 1.26~{\rm K} \left((1+z)~\frac{{\text{1 GHz}}}{\nu _{21}}\right)^{2.6}
\label{Tsystem}
\ee
Putting this together we arrive  at an expression for the noise as a function
of redshift assuming that an array is simply scaled to maintain constant
angular resolution:
\be
T_n=
\frac{12 ~\text{mK}}{\sqrt{(\tau/10^4{\rm hr}) (A_e / {\rm km^2})}}
\Bigl(\frac{1+z}{21}\Bigr)^{3.85}
\Bigl[\frac{\sqrt{1+z}}{\sqrt{1+z}-1}\Bigr]^{1/2}
\ee

The noise is a steeply rising function of redshift, but the signal
is also a strong function of redshift in the dark ages just before
reionization. For example, if star formation has not started in earnest
by $z\sim 17$,
we can see from the table below that the mean 21cm brightness temperature
at $z=30$ is 30 times larger, while the noise is higher by only a factor
of 8. The growth of fluctuations boosts the lower $z$ signal by a factor
of 1.7, but that still leaves more than a factor of 2 higher sensitivity
at $z\sim 30$ compared to $z \sim 17$. 

For both our cases we will consider 10 000 hours of total observing time.
For case I, the SKA, we consider a line of sight depth of redshift $z=15$ to 20, i.e. 374 Mpc/$h$. The effective antenna area is 1 sq km. This gives an average noise per pixel of about $T_{n}=8$~mK. The total area of sky observed is taken to be 250 degrees. 

For case II, the FFTT (or ``Omniscope''), we consider a line of sight depth of $z=25$ to 35, i.e 346 Mpc/$h$.  The effective antenna area is 100 sq km, the limit at which
earth curvature could be a problem \cite{Tegmark:2008au}. 
This gives an average noise per pixel of about $T_n=6$~mK. 
The total area of sky observed is taken to be 3000 degrees, to account for
edge effects not allowing a full 1/2 sky. 

These two cases correspond to a central redshift values of $z=17.5$ and $z=30$ respectively. Because of our finite box size we impose a lower limit in the $dq$ integration corresponding to $2\pi$ divided by the smallest dimension of the box, which is the line of sight direction. We have for the $z\in [15,20]$ and $z\in [25,30]$ a lower momentum cutoff of 0.0168~$h$/Mpc and 0.0181~$h$/Mpc, respectively. In the numerical evaluation of the integral we have checked that replacing this lower momentum cutoff by zero does not significantly affect the result.

We take our beam width $\sigma_B$ to correspond to the length scale of 1~arcminute at the comoving distance of interest. For z=17.5 and z=30, the comoving distance is 9020~Mpc/$h$ and 9650~Mpc/$h$, respectively, and the beam width is 2.63~Mpc/$h$ and 2.81~Mpc/$h$, respectively.

We use the linear matter power spectrum as provided in the LAMBDA CAMB Web Interface Toolbox~\cite{lambdacamb}. 
The power spectra can be seen in Figure \ref{fig:pk} for the different
redshift choices, along with the noise curves for the assumed experimental
parameters. For the high redshift case, the signal to noise is much higher for
two reasons: the assumed noise level is slightly smaller in power (the
increased noise at higher $z$ is assumed to be more than
offset by a larger collecting area),
and the expected cosmological mean signal (in mK) is 
substantially larger, leading to a significantly stronger constraint
on the matter power spectrum for the nominal higher redshift experiment.

We summarize our results for the two cases in the following table.

\begin{tabular}{|c|c|c|c|c|c|c|c|c|}
\hline
Case & $z$ &$\overline{\delta T_b}$ &~$T_{n}$/pixel & beam width & angular size& vol& I&$\sigma[g_{2M}]=F^{-1/2}$ \\

 & & [mK] & [mK] & $[{\rm Mpc}/h]$ & [sq. deg.] & [$({\rm Gpc}/h)^3$] &  [$(h /{\rm Mpc})^{3}$] &  \\
\hline
I & 15-20 &$-0.150$ & 8 & 2.63  & 250 &2.32 & $4.11 \times 10^{-11}$ & 4.6\\
\hline
II & 25-35 &$-4.42$ &  6 & 2.81 & 3000 &29.45  &  $2.04\times 10^{-6}$ & 0.006  \\ 
\hline
\end{tabular}

\section{Conclusions and Discussion}

The results in the above table mean that 21 cm surveys should be 
able to constrain the values of $g_{2M}$ to about
$1/\sqrt{0.0476}=4.6$ for case I and $1/\sqrt{30000}=0.006$ for 
case II.  If instead of white noise, we allow for a $P_n(q)\sim 1/q$, 
then the constraints on the values of $g_{2M}$ are 43 and 0.03 for cases I and case II, respectively. 

While the noise per mode is quite high, the
large volume allows a huge number of modes to be measured. For example, 
1 (Gpc/$h)^3$ of volume provides (in principle) more than $3 \times 10^7$ 
measurements with $\sim$ Mpc/$h$ resolution.  Even with such a large number of 
samples, we see that this is a challenging measurement
with planned upcoming experiments such as SKA; to accurately measure this signal with 21cm
experiments will necessitate deeper maps, to push the signal to noise per 
mode high enough to reach the sample variance limit. 

One way to increase the signal to noise is to measure the signal at
higher redshift, where the pre-reionization hydrogen spin 
temperature difference from the CMB is larger. At $z\sim 28$ the mean
temperature difference is 30x larger than at $z \sim 17$, in the absence
of reionization effects, while the sky noise is higher by less than a factor of 10. 
It has been suggested that reionization could
boost the $z\sim 17$ signal by up to two orders of magnitude 
\citep{Furlanetto:2006};
if the reionization process doesn't compromise the signal of interest,
this would allow SKA to have a measurement of the anisotropy to
better than ten percent.

The promise of future 21cm experiments is evident from the large signal to
noise that is possible from a sufficiently sensitive experiment. This is
in contrast to CMB experiments, where cosmological information is now
largely limited by the finite number of modes that can be measured within
the boundaries imposed by causality. For
future 21cm experiments, limits will be set by experimental capabilities,
but precise measurements at high redshift (i.e., lower frequencies) will
allow powerful constraints on fundamental physics.

\begin{acknowledgments} 
This work is supported by the FQRNT Programme de recherche 
pour les enseignants de coll\`ege, the Canadian Institute for Advanced
Research, the NSERC Discovery program, and the Canada Research Chairs
program.
\end{acknowledgments}

\end{document}